\documentclass[amsmath,amssymb,prd,a4paper,onecolumn,preprint]{revtex4} 
\usepackage[utf8]{inputenc} %
\usepackage{verbatim}    
\usepackage{graphicx}
\usepackage{dcolumn}
\usepackage{bm}
\usepackage{epsfig}
\usepackage{braket}
\usepackage{epstopdf}
\usepackage{amsfonts}
\usepackage{amssymb,amscd}
\usepackage{color} 
\usepackage[normalem]{ulem} 
\usepackage{longtable} 
\newcommand{\be}{\begin{equation}}
\newcommand{\ee}{\end{equation}}
\newcommand{\ben}{\begin{eqnarray}}
\newcommand{\een}{\end{eqnarray}}

\newcommand{\mat}[1]{\mbox{\boldmath{$#1$}}}

\begin{document}

\title{Revisiting the tensor $J^{PC} = 2^{--}$ meson spectrum}

\author{L.~M.~Abreu}   
\email[]{luciano.abreu@ufba.br}
\affiliation{Instituto de F\'isica, Universidade Federal da Bahia, Campus Universit\'ario de Ondina, Salvador, Bahia, 40170-115, Brazil}

\author{ F.~M.~da~Costa~J\'unior}
\email[]{francisco.miguel@ifsertao-pe.edu.br}
\affiliation{Instituto de F\'isica, Universidade Federal da Bahia, Campus Universit\'ario de Ondina, Salvador, Bahia, 40170-115, Brazil}
\affiliation{Instituto Federal do Sert\~ao Pernambucano, Petrolina, Pernambuco, Brazil}

\author{ A.~G.~Favero}
\email[]{aline.favero@mail.mcgill.ca}
\affiliation{Department of Physics, McGill University, Montr\'eal, QC, H3A 2T8, Canada}

\begin{abstract}

This work is devoted to the discussion and characterization of the tensor $2^{-(-)}$ meson spectrum, by making use of the Coulomb gauge Hamiltonian approach to QCD, with the interactions being given by an improved confining potential and a transverse hyperfine interaction, whose kernel is a Yukawa-type potential. Our aim is to study the basic features of $2^{-(-)}$ mesons within an unified framework through the whole range of quark masses. We concentrate our investigation on predictions of expected but yet-unobserved ground states of unflavored light mesons and on charmonium and bottomonium states. The numerical results are compared with existing literature.
\end{abstract}

\maketitle

\section{Introduction}
\label{Introduction}

Thanks to the joint efforts of experimentalists and theorists, the study of hadron spectrum has experienced a great advancement over the last decades~\cite{Tanabashi:2018oca}. But despite this tremendous progress, both light hadron and heavy quarkonium spectroscopies remain at the forefront of particle physics. In the context of charmonium and bottomonium sectors, for example, beyond the fact that there are several states predicted by quark models but not yet observed experimentally, different facilites (BELLE, BABAR, CLEO, BESIII, LHCb, etc.) have discovered new hadrons that do not exhibit the expected properties of conventional hadrons~\cite{Tanabashi:2018oca}. These states are known as $X Y Z$ states and some of them, like the charged states, are unequivocally exotic~\cite{Tanabashi:2018oca,Brambilla:2010cs,Esposito:2014rxa,Brambilla:2019esw}. The underlying structures of these new states are still in debate~\cite{Brambilla:2019esw}. 

Looking at the light hadron spectrum sector, light meson families continue attracting great physical interest due to some of their fundamental and still unclear aspects. From the experimental perspective, this importance can be attested by the running experiments like BESIII and COMPASS and forthcoming experiments like GlueX and PANDA, also dedicated to the analysis of the properties of new lighter-mass mesons and not-yet observed hybrid mesons. 
On theoretical grounds, recently a lot of work has been consecrated to establish the light meson spectrum as well as to better understand new light hadron states; see for example Refs.~\cite{Ebert:2009ub,Chen:2011qu,Pang:2014laa,Wang:2014sea,Chen:2015iqa,Pang:2015eha,Koenigstein:2016tjw,Piotrowska:2017rgt,Wang:2017iai,Pang:2017dlw,Giacosa:2017pos,Guo:2019wpx,Wang:2019qyy}.

Focusing especially on meson families of spectrum with quantum numbers $J^{P(C)} = 2^{-(-)} $, it can be identified intriguing features. First, from PDG~\cite{Tanabashi:2018oca} we notice that only four lowest-lying ground mesons have been confirmed: the strange mesons $K_2(1770)$ and $K_2(1820)$, the charmonium $\psi_2(3823)$ and the bottomonium $\Upsilon_2(1D)$. In the case of unflavored light meson families $\rho_2 ,\,  \omega_2, \, \phi_2$, the four observed states $\rho_2(1940) , \,  \omega_2(1975), \,  \omega_2(2195) , \, \rho_2(2225)$ are interpreted as ``further states'', since the ground states should be expected in the same region of mass spectrum of their $1^{--}$ and $3^{--}$ partners, i.e. $\simeq 1700$ MeV, as remarked for instance in Refs.~\cite{Godfrey:1985xj,Godfrey:1998pd,Chen:2011qu,Guo:2019wpx}.

It should be observed that the cited works make use of a variety of approaches for evaluating $2^{-(-)}$ meson spectrum; for example the relativized quark model~\cite{Godfrey:1985xj},  QCD sum rule analysis~\cite{Chen:2011qu,Sungu:2020azn}, relativized quark model for the ground states and Regge Phenomenology for excited states of light mesons~\cite{Guo:2019wpx}, and so on. Some of them report predictions for meson masses that differ up to hundreds of MeV.

Thus, this work intends to contribute to the discussion and characterization of the tensor $2^{-(-)}$ meson spectrum, by making use of a different formalism with respect to the preceding analyses. In this sense, we employ the Coulomb gauge Hamiltonian approach to QCD~\cite{Christ:1980ku,Szczepaniak:2001rg,LlanesEstrada:2001kr,Ligterink:2003hd,LlanesEstrada:2004wr,Abreu:2019adi}, with the assumption that the interactions between quarks and antiquarks are given by an improved confining potential and a transverse hyperfine interaction, whose kernel is a Yukawa-type potential. Our aim is to study the basic features of $2^{-(-)}$ mesons within an unified scheme. We perform a comparison of our results with other works. We concentrate our investigation on the unflavored light mesons and the relation between the ground states and radial excited states of charmonia and bottomonia.

The paper is organized as follows. In Section~\ref{Formalism}, we present the Coulomb–gauge QCD model within Tamm-Dancoff approximation adapted to the context of $2^{-(-)}$ states. Section~\ref{Results} is devoted to show the numerical calculations of the mass spectrum as well as the Regge trajectories in $(n,M^2)$ plane for lowest-lying and radially excited charmonia and bottomonia. Concluding remarks are in Section~\ref{Conclusions}.

\section{The formalism}
\label{Formalism}


Noticing that the interest of this work is focused on the spectrum of $q\bar{q}$ states, our starting point is an effective version of the Coulomb gauge QCD Hamiltonian~\cite{Christ:1980ku,Szczepaniak:2001rg,LlanesEstrada:2001kr,Ligterink:2003hd,LlanesEstrada:2004wr,Abreu:2019adi}. It is achieved by excluding pure gluonic contributions and employing a phenomenological approach to the quark sector, and may be written as
\begin{equation}
	H_{QCD} = H_{q} + H_{C} + H_{T}, 
\label{H_QCD1}
\end{equation}
where 
\begin{eqnarray}
	H_{q} & = & \int d\mathbf{x} \Psi^{\dagger}\left(\mathbf{x}\right) \left[-i \mat{\alpha} \cdot \mat{\nabla} + \beta m\right] \Psi\left(\mathbf{x}\right), 
	\nonumber \\
	H_{C} & = &  -\frac{1}{2} \int d\mathbf{x} d\mathbf{y} \rho^{a}\left(\mathbf{x}\right) \hat{V}\left(\vert\mathbf{x} - \mathbf{y}\vert\right) \rho^{a}\left(\mathbf{y}\right),	\nonumber \\
    H_{T} & = & \frac{1}{2} \int d\mathbf{x} \:d\mathbf{y} J_{i}^{a}\left(\textbf{x}\right) \hat{U}_{ij}\left(\mathbf{x}, \mathbf{y}\right) J^{a}_{j}(\textbf{y}). 
	\label{H_QCD2}
\end{eqnarray}
In equations above, $\Psi$ and $m$ are the current quark field and mass; the color densities $\rho^{a}$ and quark color currents $\mathbf{J}^{a}$ are given by 
\begin{eqnarray}
  	\rho^{a}(\textbf{x}) & = & \Psi^{\dagger}\left(\mathbf{x}\right) T^{a} \Psi\left(\mathbf{x}\right),  \nonumber \\
	\mathbf{J}^{a} & = & \Psi^{\dagger}\left(\mathbf{x}\right) \mat{\alpha} T^{a} \Psi\left(\mathbf{x}\right), 
	\label{color_dens_curr}
\end{eqnarray}
with $T^{a} = \lambda / 2$ and $f^{abc}$ ($a=1,2,\ldots,8$) being the $SU_{c}(3)$ generators and structure constants, respectively. For the sake of simpler notation, the flavor indices are not explicitly displayed.  

Concerning the effective couplings, we adopt for the Coulomb longitudinal interaction $H_C$ a modified confining potential based on Yang-Mills dynamics, represented in momentum space as 
\begin{eqnarray}
	V \left( p \right) = -  \frac{12.25}{p^2}
\begin{cases} 
	\left(-12.25 \frac{ m_g^{1.93}}{p^{3.93}} \right), & \mbox{for } p < m_g, \\ 
-\frac{8.07}{p^2} \frac{\ln{\left( \frac{p^2}{m_g ^2} + 0.82 \right)^{-0.62}}}{\ln{\left( \frac{p^2}{m_g ^2} + 1.41 \right)^{0.8}}}, & \mbox{for }  p > m_g.  \end{cases}
	  	\label{Coul_Pot2}
\end{eqnarray}
%
%
%
where the parameter $m_g$ is a dynamical mass scale for the quasigluons (constituent gluons), and it is set between $500$ and $800 $ MeV. It is worthy to highlight that this potential  has been proposed in Ref.~\cite{Szczepaniak:2001rg}, through the use of a self-consistent treatment to construct the quasiparticle structure of the vacuum and determine the effective instantaneous interaction. After numerically Fourier transformed to configuration space, the resulting interaction  provides a renormalization improved short ranged behavior and long-ranged confinement,  being very nearly linear for large $r$, in reasonable agreement with the lattice calculations.

We notice that the piece $H_{T}$ is associated to the quark hyperfine interaction with the form $\vec{\alpha}\cdot\vec{\alpha}$, generated perturbatively from the second-order coupling between quarks and transverse gluons after integrating out gluonic degrees of freedom. Then, we approximate it to the effective transverse hyperfine potential with the kernel $\hat{U}_{ij}$ keeping the structure of transverse gauge condition, 
\begin{eqnarray}
\hat{U}_{ij} \left(\mathbf{x}, \mathbf{y}\right) = \left(\delta_{ij} - \frac{\nabla_{i} \nabla_{j}}{\mat{\nabla}^{2}}\right)_{\mathbf{x}} \hat{U}\left(\vert \mathbf{x} - \mathbf{y} \vert\right).
\label{U_ij}
\end{eqnarray}
The form of $\hat{U}$ is chosen to mimic one-gluon exchange potential; it is given by a Yukawa-type potential, 
\begin{eqnarray} 
	U \left( p \right) = C_h\begin{cases} 
	(- 24.57) \frac{1}{p^2 + m_g ^2}, & \mbox{for } p < m_g, \\ 
- \frac{8.07}{p^2} \frac{\ln{\left( \frac{p^2}{m_g ^2} + 0.82 \right)^{-0.62}}}{\ln{\left( \frac{p^2}{m_g ^2} + 1.41 \right)^{0.8}}}, & \mbox{for }  p > m_g.  \end{cases}
	  	\label{Yuk_pot}
\end{eqnarray}
The constant $C_h$ is coded as a global strength, and the factor $-24.57$ is determined by matching the high and low momentum ranges at the scale $m_g$.

The quark gap equation is yielded following the standard Bogoliubov-Valatin (BV) variational method, via the minimization of vacuum expectation value of the Hamiltonian with respect to the quasiparticle vacuum~\cite{LlanesEstrada:2001kr,Ligterink:2003hd,LlanesEstrada:2004wr,Abreu:2019adi}: 
\ben
k s_k - m_f c_k & = & \int_{0} ^{\infty} \frac{q^2}{6 \pi ^2} \left[ s_k c_q \left(V_1 + 2 W_0 \right) - s_q c_k \left(V_0 + U_0 \right)\right]  , 
\label{gap_eq}
\een
where the functions $s_k \equiv \sin{\phi_k} $ and $c_k \equiv \cos{\phi_k}$ are defined in terms of the Bogoliubov
angle $\phi_k$ and are related to the running quark mass $M_q (k)$ through the relationship $M_q(k) = k \tan{\phi_k} $. The functions $V_0, V_1 $ and $U_0$ are associated to the longitudinal and transverse potentials and denote angular integrals in the form
\be
F_n (k,q) \equiv \int_{-1} ^{1} dx \; x^n \; F(|\mathbf{k} - \mathbf{q}|), 
\label{ang_int}
\ee
 with $x = \hat{k}\cdot \hat{q}$.  The $W$-function is given by
\be 
W(|\mathbf{k} - \mathbf{q}|) \equiv U(|\mathbf{k} - \mathbf{q}|) \frac{x (k^2 + q^2) - k q (1 + x^2) }{|\mathbf{k} - \mathbf{q}|^2}. 
\label{W_func}
\ee

After obtaining the explicit expressions for the constituent quark interaction and the dynamical quark mass, we are able to analyze mesonic bound states. In this sense,  working in the context of Tamm-Dancoff (TDA) approximation, which appears to be suitable for a wide range of meson types (except for the pion~\cite{LlanesEstrada:2001kr,Abreu:2019adi}), the equation of motion for a open-flavor meson is given by
\begin{eqnarray}
  	\langle \Psi^{nJP} \vert \left[H, Q^{\dagger}_{nJP} \right] \vert \Omega \rangle = \left(E_{nJP} - E_{0}\right) \langle \Psi^{nJP} \vert Q^{\dagger}_{nJP} \vert \Omega \rangle, 
	\label{TDA_eq}
\end{eqnarray}
where  $| \Psi ^{nJP} \rangle $ means an open-flavor meson state with total angular momentum $J$, parity $P$ and radial quantum number $n$; $Q^{\dagger}_{nJP}$ is the mesonic creation operator, given by
\begin{eqnarray}
	Q^{\dagger}_{nJP} = \sum_{\alpha \beta} \int\frac{d\mathbf{k}}{\left(2\pi\right)^{3}} \Psi^{nJP}_{\alpha \beta}\left(\mathbf{k}\right) B^{\dagger}_{\alpha}\left(\mathbf{k}\right) D^{\dagger}_{\beta}\left(-\mathbf{k}\right),
\label{meson_op}
\end{eqnarray}
with $B^{\dagger}_{\alpha}$ and $ D^{\dagger}_{\beta}$ being the quasiparticle operators, $\alpha, \beta $ denoting helicities (we have omitted the color indices), and $\Psi^{nJP}_{\alpha \beta}$ the corresponding wave function.

The TDA equation of motion in Eq.~(\ref{TDA_eq})  can be expressed in a more tractable form, by evaluating the commutators in the left-hand side after normal ordering with respect to the BCS vacuum, and also expanding the wave function in partial-waves. The final expression is written as
\begin{eqnarray}
 \left(M_{nJP} - \epsilon_{k}^{f} - \epsilon_{k}^{f'} \right) \Psi^{nJP}_{LS}\left(k\right) = \sum_{\Lambda \Sigma} \int\limits_{0}^{\infty} \frac{q^{2} dq}{12 \pi^{2}} \; K^{JP; f f'}_{L S ; \Lambda \Sigma}\left(k, q\right) \Psi^{nJP}_{\Lambda \Sigma}\left(q\right), 
\label{TDA_eq_part_wav}
\end{eqnarray}
where $\mathbf{L}$ and $\mathbf{S}$ are the orbital and spin angular momenta, respectively; $\Psi^{nJP}_{LS}\left(k\right)$ is the radial wave function; $M_{nJP} \equiv E_{nJP} - E_0$ is the mass of the meson state; $\epsilon_{k } ^{f} $ is the self-energy of the quasiparticle with flavor $f$, 
\be 
\epsilon _{k } ^{f} = m _{f} s_{k} ^{f} + k c_{k} ^{f} - \int_{0} ^{\infty} \frac{q^2}{6 \pi ^2} \left[ s_{k }^{f} s_{q}^{f} \left(V_0 + 2 U_0 \right) + c_{k}^{f} c_{q}^{f} \left(V_1 + W_0 \right)\right];
\label{self_en}
\ee
and $K^{JP; f f'}_{L S ; \Lambda \Sigma}\left(k, q\right)$ is the kernel coupling different orbital and spin states. 

We remind the reader that in Ref.~\cite{Abreu:2019adi} the kernel $K^{JP; f f'}_{L S ; \Lambda \Sigma}$ is given in ${\bf L}$--${\bf S}$ basis, with the specific expressions for the pseudoescalar ($0^{-+}$), vector ($1^{--}$) and  axial ($1^{+ \pm }$) meson states written down explicitly. But keeping in mind that the case of interest is the $2^{-(-)}$ meson state, we can use the general formula for $K^{JP; f f'}_{L S ; \Lambda \Sigma}$ to express the kernel  $K^{(2^{--})}$ in the following form
(taking into account only the lowest orbital partial-wave component and neglecting all coupling to the gluon sector) 
\begin{eqnarray}
			K^{(2^{--})}\left(k, q\right)  & = & k q Z_1 \left(a_{1} + a_{2}\right) 
			+ \frac{1}{2} \left( 2 Z_{2} - Z_0 \right) \left(a_{3} + a_{4}\right)
\nonumber  \\
			& & +   V_{3} \left(a_{5} + a_{6}\right) +\frac{1}{2} \left( 3 V_{2} - V_{0}\right) \left(a_{7} + a_{8}\right),  
	\label{K_T3}
\end{eqnarray}
where the coefficients $a_i$ are given by
\begin{eqnarray}
	a_{1} &= & \sqrt{1 + s_{k}^f } \sqrt{1 + s_{k}^{f'} } \sqrt{1 - s_{q}^f } \sqrt{1 - s_{q}^{f'} },\label{eq:d17} \nonumber \\
	a_{2} &= & \sqrt{1 - s_{k}^f } \sqrt{1 - s_{k}^{f'} } \sqrt{1 + s_{q}^f } \sqrt{1 + s_{q}^{f'} }, \nonumber \\
	a_{3} &= &\sqrt{1 + s_{k}^f } \sqrt{1 - s_{k}^{f'} } \sqrt{1 - s_{q}^f } \sqrt{1 + s_{q}^{f'} }, \nonumber \\
	a_{4} &= & \sqrt{1 - s_{k}^f} \sqrt{1 + s_{k}^{f'} } \sqrt{1 + s_{q}^f} \sqrt{1 - s_{q}^{f'} }, \nonumber \\
	a_{5} &= & \sqrt{1 + s_{k}^f } \sqrt{1 - s_{k}^{f'} } \sqrt{1 + s_{q}^f } \sqrt{1 - s_{q}^{f'} }, \nonumber \\
	a_{6} &= & \sqrt{1 - s_{k}^f } \sqrt{1 + s_{k}^{f'} } \sqrt{1 - s_{q}^f } \sqrt{1 + s_{q}^{f'} }, \nonumber \\
	a_{7} &= & \sqrt{1 + s_{k}^f } \sqrt{1 +  s_{k}^{f'} } \sqrt{1 + s_{q}^f } \sqrt{1 + s_{q}^{f'} }, \nonumber \\
	a_{8} &= & \sqrt{1 - s_{k}^f } \sqrt{1 - s_{k}^{f'} } \sqrt{1 - s_{q}^f } \sqrt{1 - s_{q}^{f'} };
	\label{a_coeff}
\end{eqnarray}
the function $s_{k(q)}^{f(f')} $ is dependent of the respective gap angle obtained by solving the gap equation for the $f(f')$-th quasiparticle. Beyond the functions $V_n$, $U_n$ and $W_n$ defined in Eqs.~(\ref{ang_int}) and ~(\ref{W_func}), we have also made use of the auxiliary $Z$-function: 
\be 
Z(|\mathbf{k} - \mathbf{q}|) \equiv U(|\mathbf{k} - \mathbf{q}|) \frac{ 1 - x^2 }{|\mathbf{k} - \mathbf{q}|^2}. 
\label{Z_func}
\ee

It should be mentioned that the TDA equation and the kernel written above have been obtained in general case of meson states with open flavor $f\neq f'$, in which the quasiparticles have different gap angles. In this context, $C$-parity is no longer a good quantum number. 
But in the case where quark and antiquark have equal flavor (i.e. states with hidden flavor  $f=f'$), the solutions of gap equation in Eq.~(\ref{gap_eq}) are the same for both. Therefore, $C$-parity becomes a good quantum number and it can be easily checked from Eq.~(\ref{a_coeff}) that the combinations of coefficients $a_i$ appearing in Eq.~(\ref{K_T3}) are simplified as follows:  $a_{1} + a_{2} = 2 (1 - s_k s_q )$, $a_{3} + a_{4} = a_{5} + a_{6} = 2 c_k c_q $, and $a_{7} + a_{8} = 2 (1 + s_k s_q )$. 

\section{Numerical Results}
\label{Results}

Here we present the results obtained by applying the TDA approach of the Coulomb–gauge QCD model, outlined in previous Section, for the $2^{-(-)}$ meson states. We solve numerically the gap equation in Eq.~(\ref{gap_eq}), which provides the gap angles to be used in the TDA equation of motion in Eq.~(\ref{TDA_eq_part_wav}). We stress that these solutions have been calculated by considering both the improved confining potential and the transverse hyperfine interaction, whose kernel is a Yukawa-type potential, being interpreted as the exchange of a constituent gluon. The input parameters of the model are then: dynamical mass of constituent gluon $m_g $, current quark masses $m_{f(f')}$ and the magnitude of transverse potential $C_h$, which in principle are chosen in order to generate findings in consonance with observed states. However, only four $2^{-(-)}$ ground states have been observed hitherto (see discussion below). 
Then, remarking that our purpose is to give the basic picture of the tensor $2^{-(-)}$ meson spectrum, and therefore access its global properties, our starting point is the set of parameters used in Ref.~\cite{Abreu:2019adi}, picked out to yield agreement with the ground states of pseudoscalar and vector mesons. Nevertheless, since this mentioned work has investigated the behavior of axial mesons with the quark mass, in the present analysis we also explore other aspects of the model, as the dependence of results on these mentioned parameters.


\begin{figure}[htbp]
\centering
\includegraphics[{width=0.6\textwidth}]{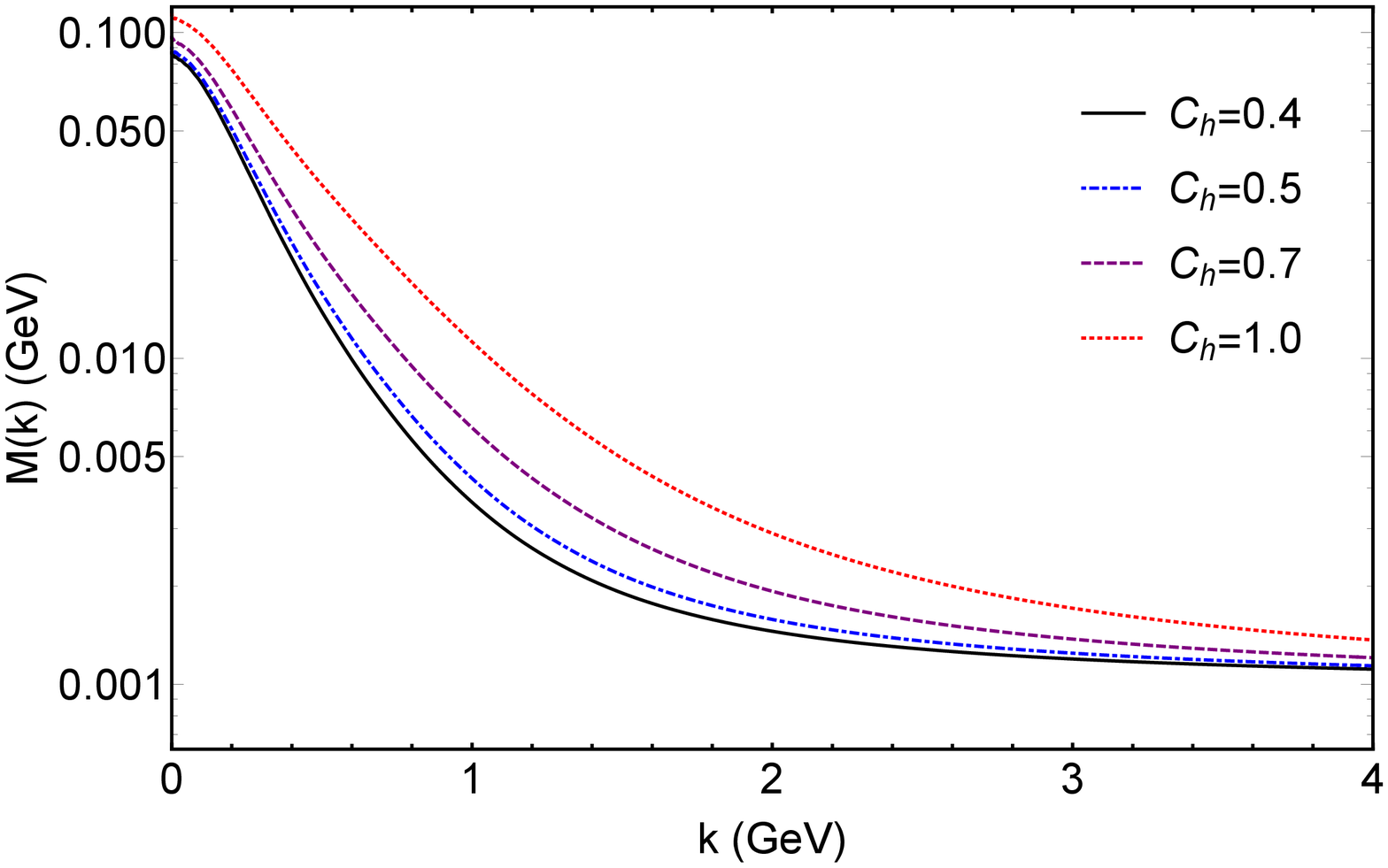} \\
\includegraphics[{width=0.6\textwidth}]{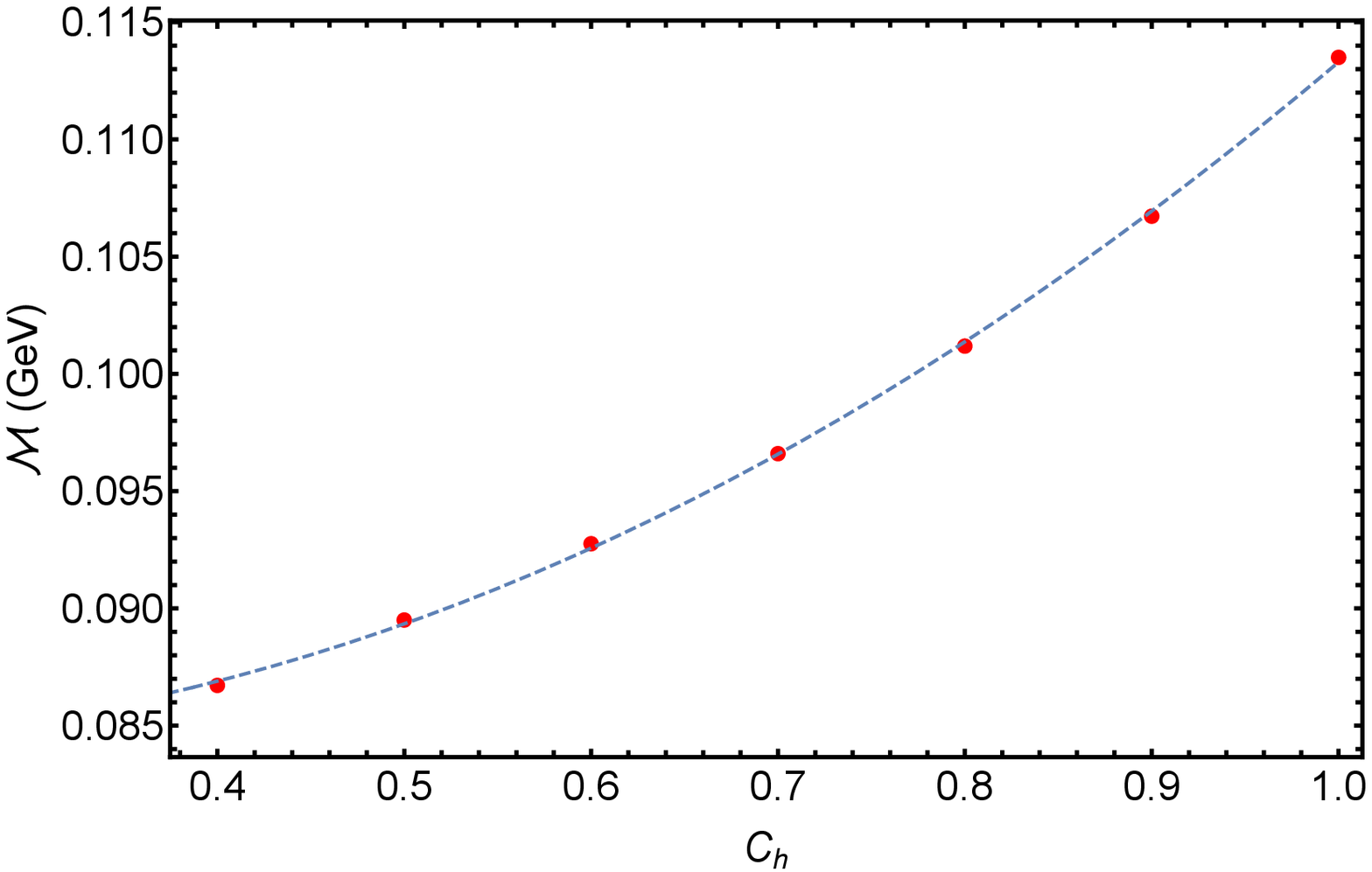}
\caption{Top panel: Running quark mass $M(k)$ as a function of momentum $k$, for different values of $C_h$. Bottom panel: constituent quark mass $\mathcal{M} = M (0) $ as a function $C_h$. The parameters used are $m_g = 600 $ MeV,  $m_f=1$ MeV. The integrations in Eq.~(\ref{gap_eq}) have been performed with a cutoff  $\Lambda = 6.0 $ GeV. }\label{fig:massgap}
\end{figure}

For completeness, we begin by showing in Fig.~\ref{fig:massgap} the behavior of running quark mass $M(k)$ (obtained from numerical solution of gap equation~(\ref{gap_eq})) with the parameter $C_h$. The constituent quark masses $\mathcal{M}$ can be extracted from the limit $k \rightarrow 0 $, i.e. $\mathcal{M} \equiv M (0) $, while at high scales the current quark mass $m_f$ is recovered. We notice that the growth of the magnitude of transverse potential modifies the value of gap angle coming from gap equation, yielding greater values of constituent quark mass. 


\subsection{Mass Spectrum}

%
The parameters of the model (constituent gluon mass $m_g $, magnitude of transverse potential $C_h$ and bare quark masses $m_f$) used as inputs in the sequence of this subsection are listed in Table~\ref{TABLE-QUARKS}. As a complement to the chosen values of $m_f$, the respective constituent quark masses $\mathcal{M}_f$ are also shown. The parameters $m_g $ and $C_h$ are kept fixed at the values that generated the outcomes of Fig.~\ref{fig:massgap}, with the evaluation of their impact on the TDA spectrum postponed to the next subsection. Another remark is that the values of $m_f$ in set I are the same as in Ref.~\cite{Abreu:2019adi}; the sets II and III have been chosen  taking different $q_s, q_c, q_b$ masses to evaluate the behavior of computed results with them, as well as to contrast with other works and available experimental data.

As pointed out in previous investigations, this approach yields smaller constituent quark masses than other static quark models. This feature can be understood as follows: the quasiparticle self–energy in Eq.~(\ref{self_en}) has the last term between brackets incorporating contributions from the potentials $V$ and $U$. Due to the attractive nature of these potentials (see Eqs.~(\ref{Coul_Pot2}) and~(\ref{Yuk_pot})), their contributions are positive, yielding greater values of the quasiparticle self-energy and thus the effective constituent quark mass as well.  As a consequence, the TDA masses obtained from the solutions of the TDA equation~(\ref{TDA_eq}) increase, which in turn demands a reduction in the bare quark masses $m_f$ (and therefore in the constituent quark masses $\mathcal{M}_f$) to reproduce the observed spectra. On the other hand, in other approaches like quark models (see for example~\cite{Godfrey:1985xj}) the one-body part of the Hamiltonian is independent of interaction potentials, and the effect reported above does not take place. Hence, keeping in mind that the effective constituent quark mass is  model-dependent, the obtention of smaller values of $\mathcal{M}_f$ than in other static constituent approaches is an expected aspect of the Coulomb Gauge QCD model.

\begin{center}
\begin{table}[h!]
\caption{Parameters of the model (constituent gluon mass $m_g $, magnitude of transverse potential $C_h$ and bare quark masses $m_f$) used as inputs in this subsection. The parameters $m_q, m_s, m_c, m_b $ are the bare quark masses associated to the $q, s, c, b$ quarks, respectively ($q \equiv u, d$ in the limit of isospin symmetry). The constituent quark masses $\mathcal{M}_f = M_f (0)$ are also shown, as a complement to the respective chosen values of $m_f$. All quantities are given in MeV, except the values for $C_h$, that are adimensional.  }
\vskip1.5mm
\label{TABLE-QUARKS}
\begin{tabular}{c|c|c|c|c}
\hline
\hline
Quantity & Set I (Ref.\cite{Abreu:2019adi})   &  Set II &  Set III  & Other estimates   \\
\hline
$m_g$                  & 600 & 600 & 600   & 500-720  ~\cite{Szczepaniak:2001rg,LlanesEstrada:2004wr}
\\
$C_h$                  & 0.7 & 0.7 & 0.7   & -
\\
$m_q \equiv m_u = m_d$ & 1   & 1   & 1     & 1.5-5.5~\cite{LlanesEstrada:2004wr,Tanabashi:2018oca}
\\
$m_s $                & 50   & 80  & 50    & 70-120~\cite{LlanesEstrada:2004wr,Tanabashi:2018oca}
\\
$m_c $                & 830 & 950 & 900    & 1000-1400~\cite{LlanesEstrada:2004wr,Tanabashi:2018oca}
\\
$m_b $                & 3900 & 4000 & 4025  & 4000-4500~\cite{LlanesEstrada:2004wr,Tanabashi:2018oca}
\\
\hline
$\mathcal{M}_q \equiv \mathcal{M}_u = \mathcal{M}_d$ & 97   &   97   &   97     &  200-340~\cite{LlanesEstrada:2004wr,Tanabashi:2018oca}
\\
$\mathcal{M}_s $ & 208        &   273          &   208     &  450-500~\cite{LlanesEstrada:2004wr,Tanabashi:2018oca}
\\
$\mathcal{M}_c $ & 1218        &   1398          &   1340   &  1500-1600~\cite{LlanesEstrada:2004wr,Tanabashi:2018oca}
\\
$\mathcal{M}_b $ & 4436        &   4667        &   4693  &  4600-5100~\cite{LlanesEstrada:2004wr,Tanabashi:2018oca}
\\
\hline
\hline
\end{tabular}
\end{table}
\end{center}


Now we summarize the calculated results of the spectra for the $2^{-(-)}$ meson states, extracted from the numerical solutions of TDA equation in Eq.~(\ref{TDA_eq}). The masses of the lowest-lying $2^{-(-)}$ states obtained for different current quark masses of sets of parameters I, II and III are displayed in Table~\ref{TABLE-MESONS-I}. For evaluation's sake, this Table also reports some results available in literature. In our estimates we consider pure $q \overline{q}$ and $s \overline{s}$ states. A comparison among the states involving $s,c,b$ quarks calculated for the different sets in this Table allows to identify a deviation between a few tens of MeV and 150 MeV. At percentage level, the indeterminacies on these computed masses are between 3 and 7 \%, with the higher fluctuation for those involving $c$ quarks. 

Taking into account that our approach aims to incorporate simultaneously light--quark and heavy--quark symmetries, we notice that our outcomes get the spectrum approximately in accordance with other works. Although the fine-tuning of spectrum is not our main focus,  it can be especially remarked from last columns of Table~\ref{TABLE-MESONS-I} that our findings well agree with the calculated spectrum by the authors of Refs.~\cite{Godfrey:1985xj,Chen:2011qu}, with typical differences between a few tens of MeV and 100-200 MeV.
Also, the dependence of the spectrum with the current quark masses is manifested when the calculated masses for the different sets are compared. In particular, set III gives computed masses with variations of tens of MeV with respect to the ones reported in Ref.~\cite{Godfrey:1985xj}, whose formalism is based on relativized quark model. The case of Ref.~\cite{Chen:2011qu}, in which the findings have been obtained within QCD sum rule analysis, has in general better agreement with our calculated masses for the set II.

\begin{widetext}
\begin{center}
\begin{table}[h!]
\caption{TDA masses of lowest-lying $2^{-(-)}$ states obtained for the sets of parameters I-III given in Table~\ref{TABLE-QUARKS}. The TDA eigenvalue problem as well as the gap equation have been solved with the presence of an improved Cornell potential and a transverse hyperfine interaction. The fourth and last columns show some results available in literature. The masses are given in GeV. Our calculated masses are rounded to 0.001 GeV. Asterisk marks ($^{\ast}$) indicate states that were extracted from the values of their respective $3^{--}$ partners in Figs. 7 and 9 of Ref.~\cite{Godfrey:1985xj}.}
\vskip1.5mm
\label{TABLE-MESONS-I}
\begin{tabular}{c | c | c | c | c | c }
\hline
\hline
Quark content  & Our Calculated Mass & Our Calculated Mass &  Our Calculated Mass & Ref.~\cite{Godfrey:1985xj}  & Ref.~\cite{Chen:2011qu}\\
$(q_f \overline{q}_{f'})$ &  Set I & Set II & Set III  &    & \\
\hline
$q\bar{q}$  & 1.732 & 1.732 & 1.732 & 1.700           &  1.780  \\
$s\bar{q}$  & 1.795 & 1.839 & 1.795 & 1.780; 1.810    &  1.850  \\
$s\bar{s}$  & 1.852 & 1.930 & 1.852 & 1.910           &  2.000  \\
$c\bar{q}$  & 2.806 & 2.945 & 2.887 & $2.830^{\ast}$  &  2.860  \\
$c\bar{s}$  & 2.838 & 3.002 & 2.919 & $2.920^{\ast}$  &  3.010  \\
$c\bar{c}$  & 3.670 & 3.927 & 3.820 & 3.840           &  3.970   \\
$b\bar{q}$  & 6.071 & 6.173 & 6.199 & $6.110^{\ast}$  &  5.660   \\
$b\bar{s}$  & 6.097 & 6.220 & 6.245 & $6.180^{\ast}$  &  6.400   \\
$b\bar{c}$  & 6.843 & 7.065 & 7.040 & $7.040^{\ast}$  &  7.080   \\
$b\bar{b}$  & 9.907 & 10.106& 10.155 & 10.150           &  10.130   \\
\hline
\hline
\end{tabular}
\end{table}
\end{center}
\end{widetext}




It is noteworthy to highlight that from experimental perspective, up to now only the following $2^{-(-)}$ ground mesons have been observed: the strange mesons $K_2(1770)$ and $K_2(1820)$, whose quantum numbers are $I(J^P) =1/2(2^{-})$ with no definite $C$-parity; the charmed meson $\psi_2(3823)$; and the bottomed meson $\Upsilon_2(1D)$. As stated before, despite the fact that the fine-tuning of spectrum is not our main goal, we stress that by choosing the appropriate set of parameters the present formalism yields outcomes in good conformity with these observed mesons. To illustrate, in Table~\ref{TABLE-MESONS-II} is shown TDA masses of lowest-lying $2^{-(-)}$ states obtained for the set of parameters III given in Table~\ref{TABLE-QUARKS}, as well as their experimental values when available in literature. 
Examining in more detail the case of strange mesons, we remark that the online version of PDG~\cite{Tanabashi:2018oca} about the $K_2(1770)$ refers to the mini-review in 2004 edition, which is based on Ref.~\cite{Aston:1993qc}. Accordingly, the $K_2(1770)$ and $K_2(1820)$ mesons are most naturally interpreted in the context of the quark model as the observed states of the mixture of the $1^1 D _2$ and $1^3 D _2$ (i.e. singlet and triplet) ground states. That being so, as in the case of axial $K_1(1270)$ and $K_1(1400)$ mesons, the singlet and triplet assignments cannot be determined, since the strange mesons are not eigenstates of charge conjugation. Thus, since in our approach only the $2^{--}$ state is computed, the obtained energy level between the observed masses of the $K_2(1770)$ and $K_2(1820)$ can be characterized as a reasonable result. This can be better understood if we assume that the computed mass of the triplet $2^{--}$ system is a bit higher than the singlet $2^{-+}$ case, then the mixing would generate respective $1 D_2 ^{\prime}$ and $1 D_2 $ mixed states higher and smaller than triplet and singlet states, producing energy levels even closer to the observed states.

\begin{widetext}
\begin{center}
\begin{table}[h!]
\caption{TDA masses of lowest-lying $2^{-(-)}$ states, obtained for the set of parameters III given in Table~\ref{TABLE-QUARKS}, and experimental values of the respective states when available in literature. The masses are given in GeV. Our calculated masses are rounded to 0.001 GeV.  A dash (–) indicates no experimental evidence. We assume that the quark composition of the mixed isoscalar states $ \omega_2$ and $ \phi_2 $ are pure $q \overline{q}$ and $s \overline{s}$, respectively. }
\vskip1.5mm
\label{TABLE-MESONS-II}
\begin{tabular}{c | c | c | c}
\hline
\hline
Quark content  & Our Calculated Mass & State & Experimental Mass \\
$(q_f \overline{q}_{f'})$ & Set III  &       & PDG ~\cite{Tanabashi:2018oca}  \\
\hline
$q\bar{q}$  & 1.732 & $\rho_2 ; \,  \omega_2$          &  - \\
$s\bar{q}$  & 1.795 & $K_2(1770); \, K_2(1820)$     & $1.773 \pm 0.008;\, 1.819 \pm 0.012$  \\
$s\bar{s}$  & 1.852 &  $\phi_2 $                       & -   \\
$c\bar{q}$  & 2.887 & $D_2 $                           & -   \\
$c\bar{s}$  & 2.919 & $D_{s2} $                        & -   \\
$c\bar{c}$  & 3.820 & $\psi_2(3823)$                   & $3.8222 \pm 0.0012$ \\
$b\bar{q}$  & 6.199 & $B_2 $                           &  -   \\
$b\bar{s}$  & 6.245 & $B_{s2} $                        &  -   \\
$b\bar{c}$  & 7.040 & $B_{c2} $                        &  -  \\
$b\bar{b}$  & 10.155 & $\Upsilon_2(1D)$                & $10.1637 \pm 0.0014$ \\
\hline
\hline
\end{tabular}
\end{table}
\end{center}
\end{widetext}

Now we turn our attention to the unflavored light meson families, namely, the isovector $\rho_2 ;$ and the isoscalars $ \,  \omega_2; \, \phi_2$. Describing more precisely the topic mentioned in Introduction, these families undergo a curious situation: according to PDG~\cite{Tanabashi:2018oca}, the observed states in this sector of spectrum are $\rho_2(1940) ; \,  \omega_2(1975); \,  \omega_2(2195) ; \, \rho_2(2225)$, and are categorized as ``further states''.  It is interesting to notice that when we consider their partners, for instance the isovector $\rho$ and isoscalar $\omega$  meson families with quantum numbers $1^{--}$ and $3^{--}$, the states $\rho (1700), \rho _3(1690)$ and $\omega(1650), \omega_3(1670)$ are in general accepted as their respective ground states. Therefore, by correspondence we might expect isoscalar and isovector $2^{- -}$ ground states in this same region of mass spectrum. For a detailed assessment of this issue, see Ref.~\cite{Guo:2019wpx}. 
The point here is that with the results from Table~\ref{TABLE-QUARKS}, the present formalism predicts ground states $(1{}^3 D_2)$ for unflavored light meson families of the order 1730 and 1850 MeV, depending on the quark content. If we assume ideally that the quark composition of the mixed isoscalar states $ \omega_2 $ and $ \phi_2 $ are $q \overline{q}$ and $s \overline{s}$, respectively, then we get $ m_{\rho_2}, m_{\omega_2} \simeq 1730$ MeV and $ m_{\phi_2} \simeq 1850$ MeV, which are some tens of MeV different from those in Refs.~\cite{Godfrey:1985xj,Chen:2011qu,Guo:2019wpx}. Thus, our results corroborate  other findings in literature about the region of mass spectrum in which the ground states 
of unflavored light meson families $\rho_2, \,  \omega_2, \, \phi_2$ should be observed. 

Hence, this effective approach with a small number of parameters allows us to reasonably reproduce the $2^{-(-)}$ mesons observed up to now; and our expectation is that the predictions sketched above give a correct description of the basic features of spectrum, to be confirmed in future.

\subsection{Dependence of radially excited states on parameters}

In this subsection we discuss numerical results for the radially excited states and their behavior with the free parameters of the formalism, in particular under the change of the magnitude of transverse potential $C_h$ and dynamical mass of constituent gluon $m_g $.

\begin{widetext}
\begin{center}
\begin{table}[h!]
\caption{TDA masses of lowest-lying and radially excited $2^{- -}$ states of $c\bar{c}$, obtained for $m_c = 950 $ MeV, taking different values of the magnitude of transverse potential $C_h$. The  masses $E_i$ are given in GeV. Our calculated masses are rounded to 0.001 GeV. }
\vskip1.5mm
\label{TABLE-MESONS-CC}
	\begin{tabular}{l|lllll}
\hline
\hline
			\multicolumn{6}{c}{$m_c = 0.950 $ GeV; $m_{g} = 0.500$ GeV}\\
\hline
		$C_h$ & $E_{1}$ & $E_{2}$ & $E_{3}$ & $E_{4}$ & $E_{5}$\\
\hline
		$0.4$ & 3.364 & 3.690 & 3.974 & 4.229 & 4.461 \\
		$0.5$ & 3.459 & 3.780 & 4.061 & 4.314 & 4.544 \\
		$0.6$ & 3.555 & 3.872 & 4.150 & 4.399 & 4.627 \\
        $0.7$ & 3.654 & 3.966 & 4.240 & 4.486 & 4.712 \\
		$0.8$ & 3.754 & 4.061 & 4.331 & 4.575 &  4.798 \\
		$0.9$ & 3.855 & 4.157 & 4.425 & 4.666 & 4.886 \\
		$1.0$ & 3.958 & 4.255 & 4.519  & 4.756 &  4.974 \\
		$1.5$ & 4.492 & 4.767 & 5.014 & 5.238 & 5.445 \\
		$2.0$ & 5.052 & 5.307 & 5.537 & 5.747 & 5.942 \\
\hline
		\multicolumn{6}{c}{ $m_c = 0.950 $ GeV; $m_{g} = 0.600$ GeV}\\
\hline
		$C_h$ & $E_{1}$ & $E_{2}$ & $E_{3}$ & $E_{4}$ & $E_{5}$\\
\hline
		$0.4$ & 3.616 & 4.021 & 4.374 & 4.688 & 4.973 \\
		$0.5$ & 3.718 & 4.117 & 4.465 & 4.776 & 5.059 \\
		$0.6$ & 3.821 & 4.214 & 4.558 & 4.866 & 5.146  \\
		$0.7$ & 3.927 & 4.314 & 4.654 & 4.958 &  5.235 \\
		$0.8$ & 4.034 & 4.416 & 4.751 & 5.052 & 5.326 \\
		$0.9$ & 4.144 & 4.520 & 4.850 & 5.148 & 5.419  \\
		$1.0$ & 4.255 & 4.625 & 4.952 & 5.246 &  5.514 \\
		$1.5$ &  4.836 & 5.177 & 5.483 &  5.760 & 6.014 \\
		$2.0$ & 5.446 & 5.762 & 6.048 & 6.308 & 6.548 \\
\hline
\end{tabular}
\end{table}
\end{center}
\end{widetext}

\begin{widetext}
\begin{center}
\begin{table}[h!]
\caption{TDA masses of lowest-lying and radially excited $2^{- -}$ states of $b\bar{b}$, obtained for $m_b = 3900 $ MeV, taking different values of the magnitude of transverse potential $C_h$. The masses $E_i$ are given in GeV. Our calculated masses are rounded to 0.001 GeV. }
\vskip1.5mm
\label{TABLE-MESONS-BB}
	\begin{tabular}{l|lllll}
\hline
\hline
			\multicolumn{6}{c}{ $m_b = 3.900 $ GeV; $m_{g} = 0.500$ GeV}\\
\hline
		$C_h$ & $E_{1}$ & $E_{2}$ & $E_{3}$ & $E_{4}$ & $E_{5}$\\
\hline
		$0.4$ & 9.241 & 9.465 & 9.660 & 9.836 & 9.997 \\
		$0.5$ & 9.378 & 9.598 & 9.791 & 9.964 & 10.124 \\
		$0.6$ &  9.514 & 9.731 & 9.922 & 10.093 & 10.251 \\
         $0.7$ & 9.651 & 9.865 & 10.053 & 10.222 & 10.378 \\
		$0.8$ & 9.788 & 9.999 & 10.184 & 10.351 & 10.505 \\
		$0.9$ & 9.924 & 10.132 & 10.315 & 10.48 & 10.632 \\
		$1.0$ & 10.061 & 10.266 & 10.446 & 10.609 & 10.759 \\
		$1.5$ & 10.744 & 10.934 & 11.101 & 11.253 & 11.393 \\
		$2.0$ & 11.421 & 11.597 & 11.752 & 11.893 & 12.023 \\
\hline
		\multicolumn{6}{c}{ $m_b = 3.900 $ GeV; $m_{g} = 0.600$ GeV}\\
\hline
		$C_h$ & $E_{1}$ & $E_{2}$ & $E_{3}$ & $E_{4}$ & $E_{5}$\\
\hline
		$0.4$ & 9.460 & 9.739 & 9.982 & 10.201 & 10.402 \\
		$0.5$ & 9.608 & 9.884 & 10.124 & 10.340 & 10.538 \\
		$0.6$ & 9.758 & 10.029 & 10.266 & 10.479 & 10.674  \\
		$0.7$ & 9.907 & 10.174 &  10.407 & 10.618 & 10.810 \\
		$0.8$ & 10.056 & 10.319 & 10.549 & 10.757 & 10.947 \\
		$0.9$ & 10.205 & 10.464 & 10.690 & 10.894 & 11.082 \\
		$1.0$ & 10.354 & 10.609 & 10.833 & 11.034 & 11.219 \\
		$1.5$ & 11.097 & 11.332 & 11.539 & 11.726 & 11.897 \\
		$2.0$ & 11.828 & 12.046 & 12.236 & 12.409 & 12.568 \\
\hline
\end{tabular}
\end{table}
\end{center}
\end{widetext}

We evaluate our predictions on the charmonia and bottomonia spectra. In this sense, Tables~\ref{TABLE-MESONS-CC} and~\ref{TABLE-MESONS-BB} show the calculated masses for radially excited states of $c\bar{c}$ and $b\bar{b}$. It can be seen that the variation of both parameters $C_h$ and $m_g $ yields noticeable different computed masses: in the region of parameter space considered, the strengthening of magnitude of transverse potential by 0.1 augments the estimates about 100-150 MeV. But this rise is slightly smaller for higher excitations. On the other hand, the growth is more pronounced for the bottomonia. Besides, looking at the change of the other relevant parameter $m_g$, we observe that the increase in the dynamical mass of constituent gluon by 0.1 engenders greater masses about a few hundreds of MeV. 

\begin{figure}[htbp]
\centering
\includegraphics[{width=0.6\textwidth}]{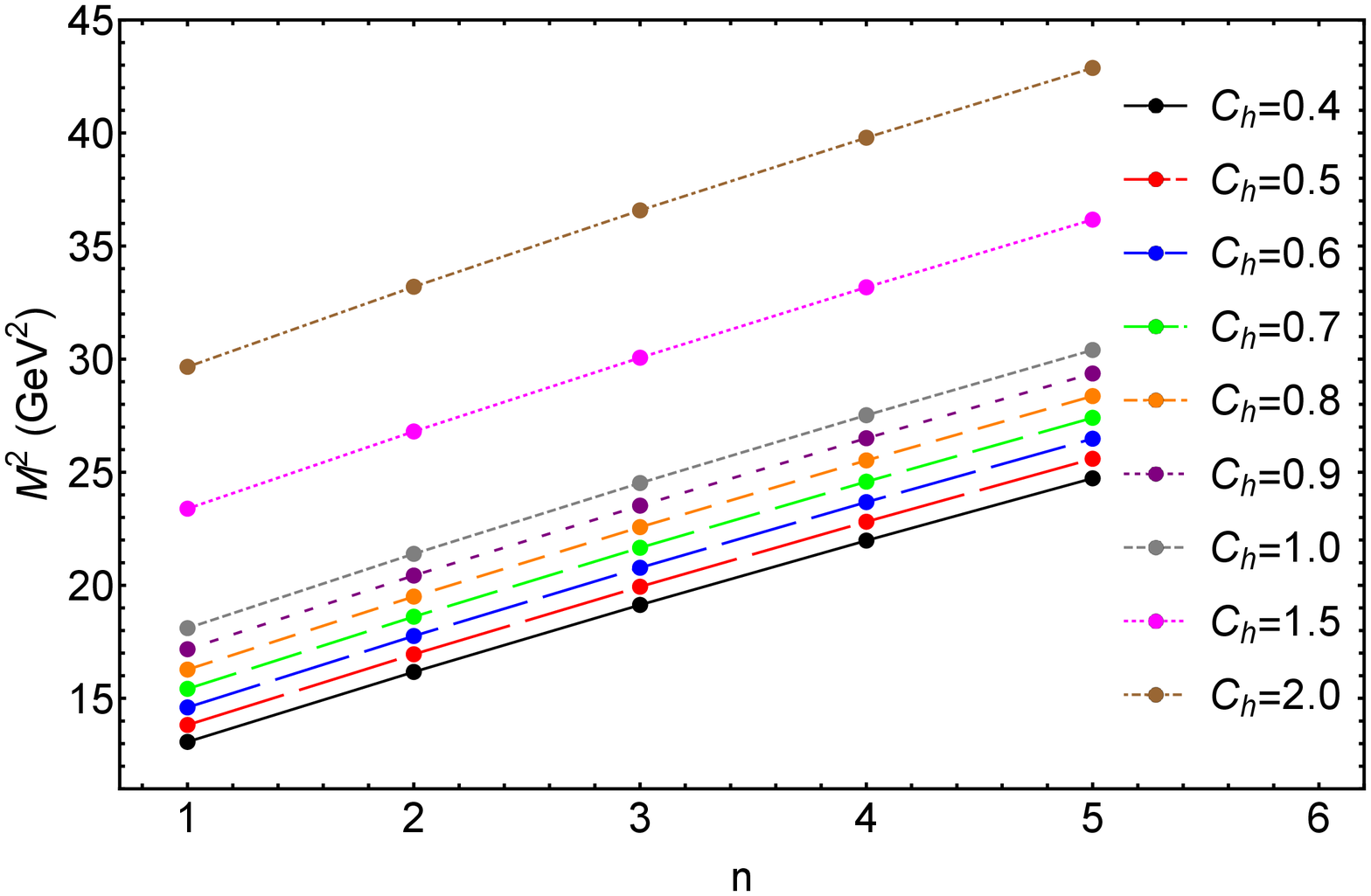} \\
\includegraphics[{width=0.6\textwidth}]{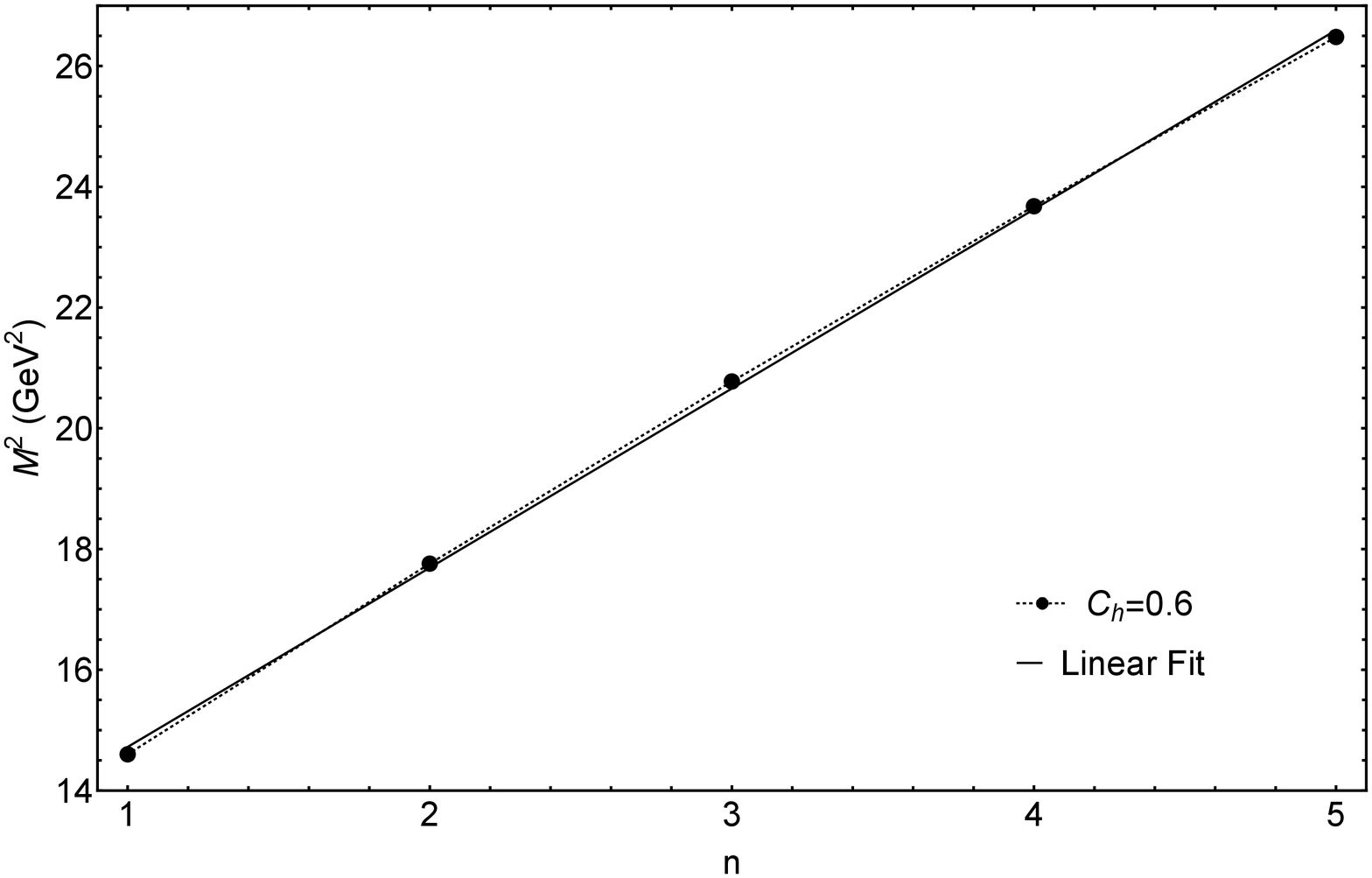}
\caption{Top panel: Regge trajectories in $(n,M^2)$ plane for $2^{- -}$ states in charmonium sector. Circles represent the predicted masses shown in Table~\ref{TABLE-MESONS-CC}, taking different values of the magnitude of transverse potential $C_h$ and at $m_g = 600 $ MeV. Bottom panel: Regge trajectory for the specific case $C_h = 0.6$ shown in Table~\ref{TABLE-MESONS-CC}, with dashed line corresponding to a nonlinear fit.}
\label{fig:Regge-Tr-cc}
\end{figure}

\begin{figure}[htbp]
\centering
\includegraphics[{width=0.6\textwidth}]{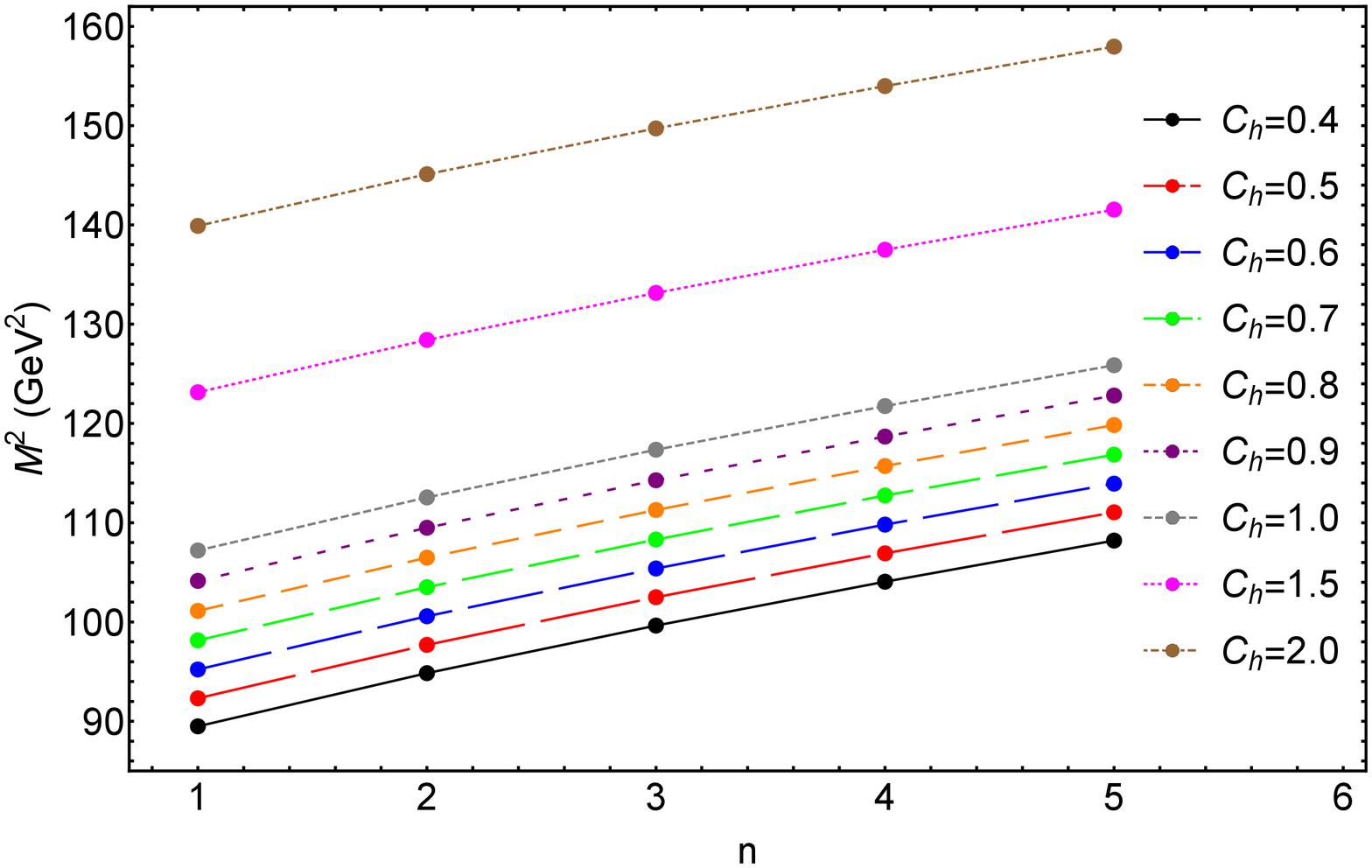} \\
\includegraphics[{width=0.6\textwidth}]{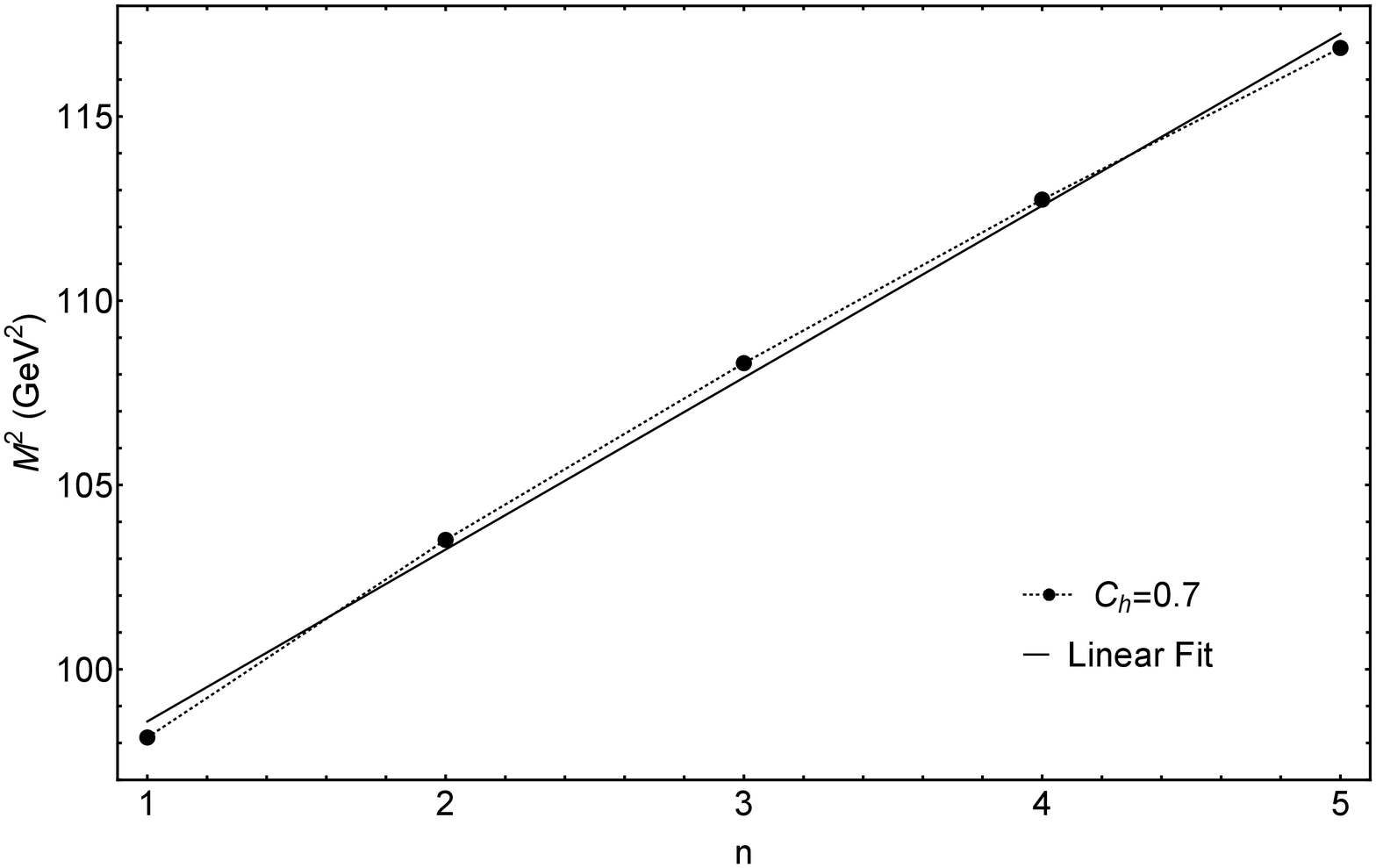}
\caption{Top panel: Regge trajectories in $(n,M^2)$ plane for $2^{- -}$ states in bottomonium sector. Circles represent the predicted masses shown in Table~\ref{TABLE-MESONS-BB}, taking different values of the magnitude of transverse potential $C_h$ and at $m_g = 600 $ MeV. Bottom panel: Regge trajectory for the specific case $C_h = 0.7$ shown in Table~\ref{TABLE-MESONS-BB}, with dashed line corresponding to a nonlinear fit.}
\label{fig:Regge-Tr-bb}
\end{figure}

Now, let us dedicate ourselves to a final remark.  As depicted in Tables~\ref{TABLE-MESONS-CC} and~\ref{TABLE-MESONS-BB}, the masses of radially excited heavy quarkonia are calculated up to high excitation number $(n = 5)$, which makes possible to obtain the mass relation between the ground states and their radial excited states, and therefore construct the Regge trajectories in the $(n,M^2)$ plane. With regard to this, in Figs.~\ref{fig:Regge-Tr-cc} and~\ref{fig:Regge-Tr-bb} are plotted these Regge trajectories for charmonia and bottomonia. It can be noticed from the results in top panels that the trajectories for different values of $C_h$ are almost parallel and equidistant, reflecting the dependence of the TDA equation on the  transverse hyperfine interaction. Furthermore, from the bottom panels it can be inferred that the behavior of mass-squared with radial quantum number is not exactly linear, which is in qualitative accordance with other works exploring different types of quarkonia states and mesons; see for instance Refs.~\cite{Ebert:2009ub,Wei:2010zza,Ebert:2011jc,Chen:2018hnx,Chen:2018bbr,Jia:2018vwl}. The charmonium case, however, exhibits  most pronounced trajectories close to linear fit. Notwithstanding, as an exercise we use the assumption that mesons are approximately grouped into radial Regge trajectories via the law form~\cite{Ebert:2009ub,Anisovich:2000kxa,Anisovich:2000ut,Afonin:2007aa}
\be
M_n ^2 = M_1 ^2 +  (n-1) \mu ^2, 
\label{LinearFit}
\ee
where $M_1$ is the mass of the lowest-lying state on each corresponding trajectory and $\mu^2$ the slope. Applying this hypothesis in our scenario, we can extract the parameter $\mu^2$ from the linear fits in specific calculations displayed in the bottom panels of Figs.~\ref{fig:Regge-Tr-cc} and~\ref{fig:Regge-Tr-bb}. We obtain the following values: $\mu^2 _{c\bar{c}} \simeq 2.9$ GeV${}^2$ and $\mu^2 _{b\bar{b}} \simeq 4.7$ GeV${}^2$ for charmonium and bottomonium, respectively. Therefore, the ratio between $\mu^2 _{c\bar{c}}$ and $\mu^2 _{b\bar{b}}$ assumes the value $ \simeq 0.6$. It gives a smaller slope for charmonium with respect to that for bottomonium. This feature is also reproduced for other heavy quarkonia states with different quantum numbers, remarking  that in our formalism the case of $2^{- -}$ mesons under analysis yields this ratio with a higher value than in other states~\cite{Wei:2010zza}.


\section{Concluding Remarks}
\label{Conclusions}

This work has been devoted to the issue of the tensor $2^{-(-)}$ meson spectrum. To this end, we have employed the Tamm-Dancoff approximation to the Coulomb–gauge QCD model by assuming that the interactions between quarks (quasiparticles) and antiquarks  (anti-quasiparticles) are given by the sum of an improved confining potential and a transverse hyperfine interaction, whose kernel is a Yukawa-type potential, being interpreted as the exchange of a constituent gluon. 

This effective approach with a small number of parameters (dynamical mass of constituent gluon $m_g $, current quark masses $m_{f}$ and the magnitude of transverse potential $C_h$) has allowed us to analyze in a global and unified framework the basic features of $2^{-(-)}$ spectrum through the whole range of quark masses. We have discussed that the calculated masses of $2^{-(-)}$ mesons can be optimized in order to fit them to the spectrum by means of fine tuning of the parameters. Besides, the estimations of expected but yet-unobserved states are approximately in accordance with other findings in literature using distinct formalisms. In particular, we contribute with predictions for the isoscalar and isovector ground states of unflavored light meson families.

Another aspect regarded has been the radially excited charmonia and bottomonia, with the analysis of the relation between the ground states and their radial excited states in the $(n,M^2)$ plane. It has been seen that the behavior of mass-squared with radial quantum number is almost but not exactly linear, which is in qualitative accordance with other works exploring other types of quarkonia states and mesons.

Some improvements can be implemented in a further work. For instance, the present analysis can be extended to incorporate the mixing of $2^{-}$ open--flavor mesons. Analogously to the case of $1^{+}$ states pointed out in Ref.\cite{Abreu:2019adi}, some mixing between the states $2^{--}$ and $2^{-+}$  is expected, yielding non-vanishing off-diagonal elements $\langle 2^{-+} \vert H \vert 2^{--} \rangle $ and $\langle 2^{--} \vert H \vert 2^{-+} \rangle $ of the Hamiltonian. In this context, $C$-parity is no longer a good quantum number. So, the TDA equation should be generalized to include non-vanishing off-diagonal elements of the Hamiltonian; and the mixing angle can be estimated on theoretical basis.

In the end, our expectation is that in the near future experimental studies of yet-unobserved $2^{-(-)}$ states will provide a solid basis for assessment of our model and our findings.

\begin{acknowledgements}

We are grateful to Felipe J. Llanes-Estrada for support and discussions. We also would like to thank the Brazilian funding agencies for their financial support: CNPq (L.M.A.: contracts 308088/2017-4 and 400546/2016-7) and FAPESB (L.M.A.: contract INT0007/2016; F.M.C.J.: contract BOL2388/2017).

\end{acknowledgements} 
%
%

\end{document}